\documentclass[a4paper,12pt]{article}

 \usepackage{amsfonts}
\usepackage{verbatim}

\begin{document}

\vskip 2cm

\begin{center}

{\bf {\Large A field-theoretic model for Hodge theory}}

\vskip 1.8cm

{\bf {\small Saurabh Gupta, R. P. Malik}}\\
{\it Centre of Advanced Studies, Physics Department,}\\
{\it Banaras Hindu University, Varanasi-221 005, (U. P.), India}\\
{\small E-mails: guptasaurabh4u@gmail.com, malik@bhu.ac.in}

\end{center}

\vskip 1.8cm

\noindent
{\bf Abstract:} We demonstrate that the four (3 + 1)-dimensional (4D)
free Abelian 2-form gauge theory presents a tractable field theoretical
model for the Hodge theory where the well-defined symmetry transformations
correspond to the de Rham cohomological operators of differential
geometry. The conserved charges, corresponding to the above continuous 
symmetry transformations, obey an algebra that is reminiscent
of the algebra obeyed by the cohomological operators. The discrete symmetry 
transformation of the theory represents the realization of the Hodge duality operation that
exists in the relationship between the exterior and co-exterior
derivatives of differential geometry. Thus, we provide the realizations
of all the mathematical quantities, associated with the de Rham cohomological
operators, in the language of the symmetries of the present 4D free
Abelian 2-form gauge theory.\\



\newpage

\noindent
{\bf {\large 1 Introduction}}\\

\noindent
The non-Abelian 1-form (i.e. $A^{(1)} = dx^\mu A_\mu$)
gauge theories, endowed with the first-class constraints
in the language of Dirac's prescription for the classification scheme [1,2], are
at the heart of the standard model of elementary particle physics where there
is a stunning degree of agreement between the theory and experiment. The above cited
first-class constraints of the 1-form gauge theories appear very naturally in the framework of 
Becchi-Rouet-Stora-Tyutin (BRST) formalism when one demands that the 
{\it true} physical states (of the total quantum Hilbert space)
are those that are annihilated (i.e. $Q_b |phys> = 0$)
by the nilpotent (i.e. $Q_b^2 = 0$) and
conserved (i.e. $\dot Q_b = 0$) BRST charge operator $Q_b$. In fact, as it turns out, it is
the operator form of the first-class constraints that annihilate the physical
states of the theory due to the physicality
condition $Q_b |phys> = 0$. This condition is consistent with
the Dirac's prescription for the quantization of systems with constraints.

The observations, made above, are true for any arbitrary $p$-form (with $p = 1, 2, 3...$)
gauge theory. The nilpotency ($Q_b^2 = 0$) of the BRST charge and the physicality
criteria ($Q_b |phys> = 0$) are the two essential ingredients
that provide a thread of connection that runs
through the cohomological aspects of  BRST charge and the exterior
derivative $d$ (with $d = dx^\mu \partial_\mu, d^2 = 0$) of
differential geometry\footnote{The exterior derivative $d = dx^\mu \partial_\mu$
(with $d^2 = 0$), the co-exterior derivative $\delta = \pm * d *$ 
(with $\delta^2 = 0$) and the Laplacian
operator $\Delta = d \delta + \delta d$ constitute the set of de Rham cohomological
operators of differential geometry on a compact manifold without a boundary.
These operators follow the algebra: $d^2 = \delta^2 = 0, \Delta = (d + \delta)^2
\equiv \{d , \delta \}, [d, \Delta ] = 0, [\delta, \Delta] = 0$. The 
operator $*$ is the Hodge duality operation on a given manifold.} [3-6].
For instance,
two BRST closed physical states (i.e. $Q_b |phys> = 0, Q_b |phys>^\prime = 0$) are
said to belong to the same cohomology class with respect to the BRST charge $Q_b$
if they differ by a BRST exact state (i.e. $|phys>^\prime = |phys> + Q_b |\chi>$
where $|\chi>$ is a non-null state). Exactly, in a similar fashion, two closed forms
(i.e. $d f = 0, d f^\prime = 0$) are said to belong to the same cohomology class w.r.t. 
$d$ if they differ by an exact form (i.e. $f^\prime = f + d g$ for $g$ to be a non-trivial 
form). Thus, the nilpotent BRST charge $Q_b$, which is the generator of the well-defined 
nilpotent BRST symmetry transformations, is a physical realization of $d$.

It has been a long-standing problem to find out the physical realizations of the other
cohomological operators $\delta = \pm * d *, \delta^2 = 0, \Delta = (d + \delta)^2$
in the language of the well-defined
symmetry properties of any arbitrary $p$-form gauge theory.
For instance, for the 4D 1-form gauge theories, the symmetry transformations were found to
be non-local and non-covariant [7-10].
In a set of papers [11-17], however, we have been 
able to obtain the well-defined symmetry transformations corresponding to
the cohomological operators
$\delta$ and $\Delta$ for the following field theoretical models, namely; 

(i)  two (1 +1)-dimensional (2D) free (non-)Abelian 1-form gauge theories
     
(without any interaction with the matter fields) [11-13],

(ii)  2D interacting U(1) Abelian gauge theory with Dirac fields [14], and

(iii) 4D free Abelian 2-form gauge theory [15-17].

In the context of the 4D Abelian 2-form gauge theory, the off-shell nilpotent
(anti-)BRST and (anti-)co-BRST symmetry transformations turn out to be anticommuting
only up to a U(1) vector gauge transformation. The absolute anticommutativity
is found to be {\it absent} in [15-17]. On the contrary, our recent work on the superfield
approach to 4D free Abelian 2-form gauge theory [18], enforces the (anti-)BRST symmetry transformations of the theory to be absolutely anticommuting due to the presence of 
a Curci-Ferrari type restriction (cf. equation (12) below) on the theory. Taking the help
of this restriction, we have been able to obtain the Lagrangian densities of the 
Abelian 2-form gauge theories that respect absolutely anticommuting (anti-)BRST
symmetry transformations [19,20]. The connection of the above type of restriction
with the concept of gerbes has also been established in  [19].

In our earlier works on the BRST approach to free 4D Abelian 2-form gauge theory [19,20],
the above 
Curci-Ferrari type restriction has been incorporated in the Lagrangian densities of the theory
through a Lorentz vector auxiliary field. As a consequence of this restriction, however, the
massless scalar field of the theory is constrained to possess a kinetic term with a negative
signature. In our very recent works [21,22], we have shown that the BRST invariant
Lagrangian densities of the 2-form gauge theory can be defined that invoke {\it no} Lagrange
multiplier (auxiliary) field. For such kind of Lagrangian densities, the Curci-Ferrari
type of restriction emerges from the equations of motion and the scalar field of the theory
is not enforced to have a negative kinetic term. These Lagrangian densities have been further
generalized [22] so as to respect the nilpotent (anti-)BRST as well as (anti-)co-BRST symmetry
transformations {\it together}. These transformations are found to be absolutely
anticommuting due to the Curci-Ferrari type restrictions.

The absolute anticommutativity of the off-shell 
nilpotent (anti-)BRST and (anti-)co-BRST symmetry transformations is very sacrosanct 
because it demonstrates the linear independence of (i) the BRST versus the anti-BRST, and
(ii) the co-BRST {\it vis-{\` a}-vis} the anti-co-BRST symmetry transformations (cf. (39) below). 
This statement
becomes crystal clear within the framework of superfield approach to BRST formalism
(see, e.g. [18]). The purpose of our present investigation is to obtain
the well-defined {\it physical} realizations of the 
{\it abstract} de Rham cohomological operators of differential geometry in the 
language of the symmetry
transformations (and the corresponding generators) in the context of 
the Abelian 2-form gauge theory. This exercise establishes that the Abelian 2-form
gauge theory is a field theoretic model for the Hodge theory where the absolute 
anticommutativity between the BRST and anti-BRST as well as co-BRST and anti-co-BRST
symmetry transformations are guaranteed due to the presence of the Curci-Ferrari type
restrictions (cf. (12) below). These restrictions emerge as the equations of motion and they
are not imposed from outside. The absolute anticommutativity is a decisive feature
of our present investigation that was absent in our earlier works [15-17].

The following central factors have motivated us to pursue our present investigation.
First and foremost, it is always very important to have a gauge theory where all
the de Rham cohomological operators find their physical meaning. Our free 4D Abelian
2-form gauge theory provides the same and, hence, it is a field theoretic model for
the Hodge theory. Second, the nilpotent (anti-)BRST and (anti-)co-BRST symmetry transformations
were {\it not} found to be absolutely anticommuting for the present model in [15-17]. It
was challenging to achieve this goal so that there could be consistency between the
superfield approach to Abelian 2-form gauge theory [18] and the ordinary field
theoretical approach to BRST formalism for the same theory. We have accomplished 
this goal in our present endeavour. Finally, our present study is a modest step
towards our main goal of studying the interacting non-Abelian 2-form gauge theories
within the framework of BRST formalism.

The outline of our present paper is as follows. In Sec. 2, we generalize
the gauge-fixed Kalb-Ramond Lagrangian density by incorporating two auxiliary vector
fields and a pair of massless scalar fields. Our Sec. 3 deals with the 
nilpotent and anticommuting
(anti-)BRST symmetry transformations and corresponding conserved charges.
Our Sec. 4 is devoted to the discussion of the anticommuting (anti-)co-BRST symmetry
transformations and the derivation of their generators. In Sec. 5, we derive
a bosonic symmetry transformation that is the analogue of the Laplacian
operator of differential geometry. Our Sec. 6 deals with the ghost and discrete
symmetry transformations of the theory. We derive the extended BRST
algebra in Sec. 7 where its relationship with the cohomological differential
operators is established at the algebraic level. Finally, in our Sec. 8, 
we make some concluding remarks
and point out a few future directions for further investigations.  

In Appendices A, B and C, we mention some key points for specific proofs.\\

\noindent
{\bf 2 Preliminaries: gauge-fixed Lagrangian densities}\\

\noindent
We begin with the Kalb-Ramond Lagrangian density [23-26] 
\begin{eqnarray}
{\cal L}^{(0)} &=& \frac {1}{12} H^{\mu \nu \kappa} H_{\mu\nu\kappa}, \qquad
H_{\mu\nu\kappa}= \partial_\mu B_{\nu\kappa}+\partial_\nu B_{\kappa\mu}+\partial_\kappa 
B_{\mu\nu},
\end{eqnarray}
where
the totally antisymmetric curvature tensor $H_{\mu \nu \kappa}$ is derived from 
the 3-form $H^{(3)}= d B^{(2)}= [(dx^\mu \wedge dx^\nu \wedge dx^\kappa)/(3!)] \;
H_{\mu\nu\kappa}$. In the above\footnote{We choose
here the flat metric (for the 4D Minkowski spacetime manifold) with signatures
(+1, -1, -1, -1). The choice of the totally antisymmetric 4D 
Levi-Civita tensor ($\varepsilon_{\mu\nu\eta\kappa}$)
is such that $\varepsilon_{0123} = + 1 = - \varepsilon^{0123}, \varepsilon_{0ijk} = \epsilon_{ijk}$,
$\varepsilon_{\mu\nu\kappa\zeta} \varepsilon^{\mu\nu\kappa\zeta} = - 4!,
\varepsilon_{\mu\nu\kappa\zeta} \varepsilon^{\mu\nu\kappa\eta} = - 3!
\delta^\eta_\zeta$, etc. Here $\epsilon_{ijk}$ is the 3D Levi-Civita tensor.
In the whole body of our text, we shall be taking
the Greek indices $\mu, \nu, \eta, \kappa....= 0, 1, 2, 3$ and the Latin indices
$i, j, k........= 1, 2, 3$.}, the 2-form 
$B^{(2)}= [(dx^\mu \wedge dx^\nu ) /(2!)] B_{\mu\nu}$ defines the antisymmetric tensor gauge potential $B_{\mu\nu}$ and $d= dx^\mu \partial_\mu $  (with $d^2=0$) is the exterior derivative of differential geometry. It can be checked 
that $\partial_\mu H^{\mu\nu\kappa} =0$ (due to the Euler-Lagrange equation of motion
derived from the above Lagrangian density).

The gauge-fixing term for the above Lagrangian density
is derived by the application of the co-exterior derivative $(\delta = - * d * , \; \delta^2= 0)$ on the 2-form  (i.e. $ \delta B^{(2)}= - * d * B^{(2)}= (\partial^\nu B_{\nu\mu}) dx^{\mu}$) which 
leads to the expression for the 1-form $G^{(1)} = (\partial^\nu B_{\nu\mu}) dx^\mu$. 
In the above, the $*$ is the  Hodge duality operation on the 4D spacetime
manifold. In fact, the explicit expression for the 
gauge-fixed Lagrangian density of the present 2-form gauge theory is 
\begin{eqnarray}
{\cal L}^{(1)} &=& \frac {1}{12} H^{\mu\nu\kappa} H_{\mu\nu\kappa} 
+ \frac {1}{2}(\partial^\nu B_{\nu\mu})(\partial_\kappa B^{\kappa\mu}).
\end{eqnarray}
The ensuing equation of motion, derived from the above gauge-fixed Lagrangian density, 
 is now  $\Box B_{\mu\nu} = 0 $ where $\Box = \partial_0^2 - \partial_i^2$ is the d'Alembertian operator. This equation of motion has its origin in the Laplacian 
operator $\Delta = (d + \delta )^2 = d \delta  + \delta  d$ when one demands the validity of the Laplace equation $\Delta B^{(2)} =0$. It is interesting to note that ${\cal L}^{(1)}$ remains invariant under the discrete symmetry 
transformation $ B_{\mu\nu}\rightarrow  \mp \frac {i}{2}  \varepsilon _{\mu\nu\eta\kappa} B^{\eta\kappa}$.

One can linearize the kinetic and gauge fixing parts of the Lagrangian density (2) by introducing the Nakanishi-Lautrup type of Lorentz vector auxiliary fields ${\cal B}^{(1)}_\mu$ 
and $B^{(1)}_\mu$ as given below:
\begin{eqnarray}
{\cal L}^{(2)} &=& \frac{1}{2} 
 {\cal B}_{\mu}^{(1)}\; {\cal B}^{\mu (1)} - {\cal B}^{\mu (1)} \;\Bigl (\frac {1}{3!} 
\varepsilon_{\mu\nu\eta\kappa} H^{\nu\eta\kappa} \Bigr ) \nonumber\\ 
&+& B^{\mu (1)} \;\Bigl (\partial^\nu B_{\nu\mu} \Bigr ) 
- \frac {1}{2}B^{\mu (1)}\; B^{(1)}_{\mu}.
\end{eqnarray}
It is worthwhile to mention that the Hodge dual of 3-form $H^{(3)}$ $(i.e. * H^{(3)} = \frac{1}{3!} (dx^{\mu}) \varepsilon _{\mu\nu\eta\kappa} H^{\nu\eta\kappa})$, which
happens to be a 1-form,
has been exploited in the linearization of the kinetic term. The above Lagrangian density 
in (3) respects the following discrete symmetry transformations: 
\begin{eqnarray}
B_{\mu\nu}\rightarrow \mp \frac{i}{2} \varepsilon_{\mu\nu\eta\kappa} B^{\eta\kappa}, 
\quad B^{(1)}_{\mu} \rightarrow \pm i {\cal B}^{(1)}_{\mu},\quad 
{\cal B}^{(1)}_{\mu} \rightarrow \mp i B^{(1)}_{\mu}.
\end{eqnarray}
The equations of motion that emerge from the above Lagrangian density are
\begin{eqnarray}
&& (\partial \cdot B^{(1)}) = 0, \qquad B^{(1)}_\mu = \partial^\nu B_{\nu\mu}, \qquad  
(\partial \cdot {\cal B}^{(1)}) = 0,  \quad \Box B^{(1)}_\mu = 0,
\nonumber\\
&& \varepsilon_{\mu\nu\kappa\eta} \partial^\kappa {\cal B}^{\eta (1)} 
+ (\partial_\mu B_\nu^{(1)} - \partial_\nu B_\mu^{(1)}) = 0, \qquad
{\cal B}^{(1)}_\mu  =  \frac {1}{3!} 
\varepsilon_{\mu\nu\eta\kappa} H^{\nu\eta\kappa},
\nonumber\\
&& \varepsilon_{\mu\nu\kappa\eta} \partial^\kappa B^{\eta (1)} 
- \bigl (\partial_\mu {\cal B}_\nu^{(1)} - \partial_\nu {\cal B}_\mu^{(1)} \bigr ) = 0, \qquad
\Box {\cal B}^{(1)}_\mu = 0, \quad \Box B_{\mu\nu} = 0.
\end{eqnarray}
The Lagrangian density (3) has room for further generalizations.

We have the freedom to add/subtract the 1-forms 
(i.e. $ F^{(1)} {(1)} = d\phi_{1} = dx^{\mu}  \partial_{\mu}\phi _{1}$ and $ F^{(1)} {(2)} = d\phi_{2} = dx^{\mu} \partial_{\mu} \phi _{2}$) to the gauge fixing term as well as the Hodge dual of the 3-form. These 1-forms are constructed with the massless
scalar fields $\phi_{1}$ and $\phi_{2}$. The above statements
can be materialized in following two different ways, namely;
\begin{eqnarray}
{\cal L}^{(3)} &=& \frac {1}{2} {\cal B}^{\mu}\; {\cal B}_{\mu} - {\cal B}^{\mu}\;
\Bigl (\frac{1}{3!} \varepsilon_{\mu\nu\eta\kappa} H^{\nu\eta\kappa} + \frac{1}{2} 
\partial_{\mu} \phi_{2} \Bigr )  \nonumber\\ &+& B^{\mu} \;\Bigl (\partial^{\nu} B_{\nu\mu} + \frac{1}{2} \partial_{\mu}\phi_{1} \Bigr ) -  \frac {1}{2} B^{\mu}\; B_{\mu},
\end{eqnarray}
\begin{eqnarray}
{\cal L}^{(4)} &=& \frac {1}{2} {\bar{\cal B}}^{\mu} \;{\bar{\cal B}}_{\mu}  - 
 {\bar{\cal B}}^{\mu} \;\Bigl (\frac{1}{3!} \varepsilon_{\mu\nu\eta\kappa} H^{\nu\eta\kappa} - \frac{1}{2} \partial_{\mu} \phi_{2} \Bigr ) \nonumber\\ &+& \bar B^{\mu} \;
\Bigl (\partial^{\nu} B_{\nu\mu}  - \frac{1}{2} \partial_{\mu}\phi_{1} \Bigr ) 
-  \frac {1}{2} {\bar B^{\mu}}\; {\bar B_{\mu}},
\end{eqnarray}
where
\begin{eqnarray}
&& {\cal B}_{\mu} = \frac{1}{3!} \varepsilon_{\mu\nu\eta\kappa} H^{\nu\eta\kappa} + \frac{1}{2} \partial_{\mu} \phi_{2}, \qquad {B}_{\mu} = \partial^{\nu} B_{\nu\mu} 
+ \frac{1}{2} \partial_{\mu}\phi_{1},  \nonumber\\
&& {\bar{\cal B}}_{\mu} = \frac{1}{3!} \varepsilon_{\mu\nu\eta\kappa} H^{\nu\eta\kappa} - \frac{1}{2} \partial_{\mu} \phi_{2}, \qquad {\bar {B}_{\mu}} = \partial^{\nu} B_{\nu\mu} - \frac{1}{2} \partial_{\mu}\phi_{1}. 
\end{eqnarray}
It should be noted that a factor of half has been taken with the scalar fields $\phi_{1}$ and $\phi_{2}$ for the algebraic convenience. The Lagrangian densities (6) and (7) are endowed with the following discrete symmetry transformations:
\begin{eqnarray}
&& B_{\mu\nu}\rightarrow \mp \frac{i}{2} \varepsilon_{\mu\nu\eta\kappa} B^{\eta\kappa}, \qquad \phi_{1}\rightarrow \pm i\phi_{2}, \qquad \phi_{2}\rightarrow \mp i\phi_{1},\nonumber\\ 
&& B_{\mu}\rightarrow \pm i{\cal B}_{\mu},\qquad {\cal B}_{\mu}\rightarrow \mp i B_{\mu},
\end{eqnarray}
\begin{eqnarray}
&& B_{\mu\nu} \rightarrow \mp \frac{i}{2} \varepsilon_{\mu\nu\eta\kappa} B^{\eta\kappa}, \qquad \phi_{1}\rightarrow \pm i\phi_{2}, \qquad \phi_{2}\rightarrow \mp i\phi_{1},\nonumber\\
&& {\bar{B_{\mu}}}\rightarrow \pm i{\bar{\cal B}}_{\mu},\qquad {\bar{\cal B}}_{\mu}\rightarrow \mp i {\bar B}_{\mu}.
\end{eqnarray}
The equations of motion (with $(\partial \cdot B) = \partial_\mu B^\mu$, etc.),
that are obeyed by the fields $ B_{\mu\nu}, \phi_{1}, \phi_{2}, B_{\mu},  {\bar B_{\mu}}, {\cal B_{\mu}}, {\bar{\cal B}_{\mu}}$ of the above Lagrangian densities, are:
\begin{eqnarray}
&& \Box {B_{\mu\nu}} = 0, \qquad \Box{\phi_{1}} = 0, \qquad \Box{\phi_{2}} = 0, 
\nonumber\\ && \Box {B_{\mu}} = 0, \quad  {\Box{\bar B_{\mu}}} = 0,  
\quad {\Box{\cal B_{\mu}}} = 0, \quad {\Box{\bar{\cal B}_{\mu}}} = 0, \nonumber\\
&& (\partial \cdot B) = 0, \quad  (\partial \cdot \bar B) = 0, \quad
   (\partial \cdot {\cal B}) = 0,
\quad (\partial \cdot \bar {\cal B}) = 0.
\end{eqnarray} 
It has been shown, in our earlier work, that the gauge-fixed Lagrangian density (2) is
endowed with a set of (dual-)gauge symmetry transformations when the (dual-)gauge parameters
of the above transformations are restricted to obey exactly similar kind of conditions
(see, e.g. [17] for details).

Before we close this section, we would like to point out that the equations in (8) imply 
the following relationships amongst the auxiliary fields, gauge fields and massless scalar fields:
\begin{eqnarray}
&& B_\mu - \bar B_\mu = \partial_\mu \phi_1,  \qquad 
{\cal B}_\mu - \bar {\cal B}_\mu = \partial_\mu \phi_2, \nonumber\\
&& B_\mu + \bar B_\mu = 2 \partial^\nu B_{\nu\mu},  \quad 
{\cal B}_\mu + \bar {\cal B}_\mu = \varepsilon_{\mu\nu\eta\kappa} \partial^\nu B^{\eta\kappa}. 
\end{eqnarray}
We note that the auxiliary fields $B_\mu, \bar B_\mu, {\cal B}_\mu, \bar {\cal B}_\mu$
as well as the massless scalar fields $\phi_1$ and $\phi_2$ obey relationships that are
reminiscent of the Curci-Ferrari restriction that exists in the realm of the non-Abelian
1-form gauge theory [27]. In the above, we have also used 
$\frac{1}{3} \varepsilon_{\mu\nu\eta\kappa} H^{\nu\eta\kappa} = \varepsilon_{\mu\nu\eta\kappa}
\partial^\nu B^{\eta\kappa}$. It is worthwhile to mention that we have already 
obtained [18] the Curci-Ferrari type restriction 
$B_\mu - \bar B_\mu = \partial_\mu \phi_1$ from the
application of the superfield approach to BRST formalism for the free 4D Abelian
2-form gauge theory. \\

\noindent
{\bf 3 Absolutely anticommuting (anti-)BRST transformations}\\

\noindent
The (anti-)BRST invariant Lagrangian densities
\begin{eqnarray}
{\cal L}_{(B,{\cal B})} &=& \frac {1}{2}{\cal B} \cdot {\cal B}  -  
{\cal B}^{\mu} \Bigl ( \frac{1}{2} \varepsilon_{\mu\nu\eta\kappa} \partial^{\nu} 
B^{\eta\kappa} +\frac{1}{2} \partial_{\mu} \phi_{2} \Bigr ) \nonumber\\
&+&  B^{\mu} \Bigl ( \partial^{\nu} B_{\nu\mu} + \frac{1}{2}\partial_{\mu}\phi_{1} \Bigr )
-\frac {1}{2} B \cdot  B + \partial_{\mu}{\bar \beta } \partial^{\mu} \beta \nonumber\\
&+& (\partial_{\mu} {\bar C}_{\nu} - \partial_{\nu}{\bar C}_{\mu})(\partial^{\mu}C^{\nu})
+(\partial \cdot C - \lambda)\rho + (\partial \cdot {\bar C}+ \rho )\lambda ,
\end{eqnarray}
\begin{eqnarray}
{\cal L}_{(\bar B, {{\bar {\cal B}})}} &=& \frac {1}{2}{{\bar{\cal B}}} \cdot 
{{\bar {\cal B}}}  - {{\bar {\cal B}}}^{\mu} \Bigl ( \frac{1}{2} \varepsilon_{\mu\nu\eta\kappa} \partial^{\nu} B^{\eta\kappa} - \frac{1}{2} \partial_{\mu} \phi_{2} \Bigr ) \nonumber\\
&+&  {\bar B}^{\mu} \Bigl ( \partial^{\nu} B_{\nu\mu} - \frac{1}{2}\partial_{\mu}\phi_{1} \Bigr )
-\frac {1}{2} {\bar B} \cdot  {\bar B} + \partial_{\mu}{\bar \beta}  \partial^{\mu} \beta \nonumber\\
&+& (\partial_{\mu} {\bar C}_{\nu} - \partial_{\nu}{\bar C}_{\mu})(\partial^{\mu}C^{\nu})
+ (\partial \cdot C - \lambda)\rho + (\partial \cdot{\bar C}+ \rho )\lambda,
\end{eqnarray}
are the generalization of the Lagrangian densities\footnote{The Lagrangian densities
in (13) and (14) are slightly different from the ones proposed in [21,22] because the former
respect the discrete symmetry transformations (9), (10) and (53) which is not the case
with the Lagrangian densities of [21,22].} 
(6) and (7) which include the bosonic (anti-)ghost fields $(\bar {\beta})\beta$, the fermionic Lorentz vector (anti-)ghost fields $(\bar C_{\mu}) C_{\mu}$ and the auxiliary $(\lambda = +\frac{1}{2}(\partial\cdot C),\; \rho = - \frac{1}{2}(\partial\cdot {\bar C}))$ (anti-)ghost fields $ (\rho)\lambda$ (with $C_\mu^2 = \bar C_\mu^2 = 0, C_\mu \bar C_\nu = - \bar C_\nu C_\mu,
\rho^2 = \lambda^2 = 0, \lambda \rho = - \rho \lambda$).

It can be checked that the following off-shell nilpotent $(s_{(a)b}^{2} = 0)$ and
anticommuting ($s_b s_{ab} + s_{ab} s_b = 0$) 
(anti-)BRST\footnote{We adopt here
the standard notations used in our earlier works [15-22].} transformations $s_{(a)b}$ :
\begin{eqnarray}
&& s_b B_{\mu\nu} = - (\partial_\mu C_\nu - \partial_\nu C_\mu), 
\qquad s_b C_\mu = - \partial_\mu \beta, \qquad s_b \bar C_\mu = - B_\mu, \nonumber\\
&& s_b \phi_{1} = -2\lambda, \qquad s_b \bar \beta = - \rho, \qquad 
s_b \bigl [\rho, \lambda, \beta, \phi_{2}, {\cal B}_{\mu}, B_\mu, H_{\mu\nu\kappa} \bigr ] = 0,
\end{eqnarray}
\begin{eqnarray}
&& s_{ab} B_{\mu\nu} = - (\partial_\mu {\bar C}_\nu - \partial_\nu {\bar C}_\mu), 
\quad s_{ab} {\bar C}_\mu = - \partial_\mu {\bar \beta}, \qquad s_{ab} C_\mu = {\bar B}_\mu, \nonumber\\
&& s_{ab} \phi_{1} = -2\rho, \qquad s_{ab} \beta = - \lambda, 
\quad s_{ab} \bigl [\rho, \lambda, {\bar\beta}, \phi_{2}, {\bar {\cal B}}_{\mu}, 
{\bar B}_\mu, H_{\mu\nu\kappa} \bigr ] = 0,
\end{eqnarray}
are the {\it symmetry} transformations for the Lagrangian densities (13) and (14), respectively, because the following relationships:
\begin{eqnarray}
s_{b}{\cal L}_{(B,{\cal B})} = -\partial_{\mu} \Bigr [(\partial^\mu C^\nu - \partial^\nu C^\mu) B_{\nu} + \rho \;\partial^{\mu}\beta +\lambda B^{\mu} \Bigl ], \nonumber\\
s_{ab}{\cal L}_{(\bar B,{\bar{\cal B}})} = -\partial_{\mu} \Bigr [(\partial^\mu {\bar C}^\nu - \partial^\nu {\bar C}^\mu) {\bar B}_{\nu} - \rho \; {\bar B}^{\mu} + \lambda \partial^{\mu} {\bar \beta} \Bigl ],
\end{eqnarray}
ensure that
(13) and (14) transform to the total spacetime derivatives.

Interestingly, the following explicit forms of (13) and (14), namely;
\begin{eqnarray}
{\cal L}_{(B,{\cal B})} &=& \frac {1}{2}{\cal B}_{\mu} \; {\cal B}^{\mu} -  
{\cal B}^{\mu} \;\Bigl ( \frac{1}{2} \varepsilon_{\mu\nu\eta\kappa} \partial^{\nu} B^{\eta\kappa} +\frac{1}{2} \partial_{\mu} \phi_{2} \Bigr ) \nonumber\\
&+& s_b \Bigr [-{\bar C}^{\mu} \{(\partial^{\nu} B_{\nu\mu} + \frac{1}{2}\partial_{\mu}\phi_{1})- \frac{1}{2} B _{\mu}\} + {\bar\beta}(\partial\cdot  C - 2\lambda ) \Bigl ],
\end{eqnarray}
\begin{eqnarray}
{\cal L}_{(\bar B,{\bar{\cal B}})} &=& 
\frac {1}{2} \bar {\cal B}_{\mu}\; \bar {\cal B}^{\mu}  -  {\bar{\cal B}}^{\mu}\;
 \Bigl ( \frac{1}{2} \varepsilon_{\mu\nu\eta\kappa} \partial^{\nu} B^{\eta\kappa} -\frac{1}{2} \partial_{\mu} \phi_{2} \Bigr ) \nonumber\\
&+& s_{ab} \Bigr [+C^{\mu} \{(\partial^{\nu} B_{\nu\mu} - \frac{1}{2}\partial_{\mu}\phi_{1})- \frac{1}{2} {\bar B _{\mu}\} + \beta(\partial\cdot {\bar C} + 2\rho) } \Bigl ],
\end{eqnarray}
demonstrate the BRST and anti-BRST invariance of (13) and (14) in a very simple manner. This is due to the nilpotency of the (anti-)BRST transformations (i.e. $s_{(a)b}^{2}= 0$ and the fact that $s _{(a)b} [{\cal B}_{\mu},{\bar{\cal B}}_{\mu}, \phi_{2},\varepsilon_{\mu\nu\eta\kappa}\partial^{\nu}B^{\eta\kappa}  ] = 0$).
It is worthwhile to point out that the (anti-)BRST symmetry transformations (15) and (16) are absolutely  anticommuting as it can be seen that 
\begin{eqnarray}
(s_b s_{ab} + s_{ab} s_b)\; B_{\mu\nu} \equiv \{ s_b, s_{ab} \}\; B_{\mu\nu} = 
\partial_{\mu}(B_\nu- \bar B_\nu)- \partial_\nu (B_{\mu}-{\bar B}_{\mu}) = 0,
\end{eqnarray}
is automatically satisfied because of the Curci-Ferrari type of restriction (12) (where
$B_{\mu} - {\bar B}_{\mu} = \partial_{\mu}\phi_{1}$). All the rest of the fields of the theory
respect the anticommuting property as can be checked explicitly by exploiting the transformations (15) and (16) supplemented with the following inputs\footnote{Exploiting the transformations
(15), (16) and (21), it can be checked that $s_{ab} {\cal L}_{(B, {\cal B})} 
= - \partial_\mu [ (\partial^\mu \bar C^\nu - \partial^\nu \bar C^\mu) B_\nu 
- \rho (\bar B^\mu + \partial_\nu B^{\nu\mu})  + \lambda \partial^\mu \bar \beta ]
+ (\partial^\mu \bar C^\nu - \partial^\nu \bar C^\mu) \partial_\mu (B_\nu - \bar B_\nu)
- (\partial^\mu \rho) (\bar B_\mu + 2 B_\mu - \frac{1}{2} \partial_\mu \phi_1)$ and
$s_{b} {\cal L}_{(\bar B, \bar {\cal B})} 
= - \partial_\mu [ (\partial^\mu  C^\nu - \partial^\nu  C^\mu) \bar B_\nu 
+ \lambda (B^\mu + \partial_\nu B^{\nu\mu})  + \rho \partial^\mu  \beta ]
- (\partial^\mu  C^\nu - \partial^\nu  C^\mu) \partial_\mu (B_\nu - \bar B_\nu)
+ (\partial^\mu \lambda) (B_\mu + 2 \bar B_\mu + \frac{1}{2} \partial_\mu \phi_1)$.
Due to the equations (12), it can be checked that
the Lagrangian densities (13) and (14) remain invariant under anti-BRST and BRST
transformations (16) and (15), respectively, on the constrained surface where
$B_\mu - \bar B_\mu = \partial_\mu \phi_1$, $\bar B_\mu + 2 B_\mu - \frac{1}{2} 
\partial_\mu \phi_1 = 3 \partial^\nu B_{\nu\mu}, 
B_\mu + 2 \bar B_\mu + \frac{1}{2} \partial_\mu \phi_1 = 3 \partial^\nu B_{\nu\mu}$.}:
\begin{eqnarray}
s_{b} {\bar B}_{\mu} = -\partial_{\mu}\lambda,\qquad s_{ab} B_\mu = \partial_{\mu}\rho,
\qquad s_b {\bar{\cal B}}_{\mu} = 0, \qquad s_{ab} {\cal  B}_{\mu} = 0.
\end{eqnarray}
The set of nilpotent transformations (15), (16) and (21) implies
that the anticommutativity property (i.e. $\{s_b,s_{ab}\} = 0$) is clearly satisfied.

Exploiting the Noether's theorem, the following conserved currents
\begin{eqnarray}
J^{\mu}_{(b)}&=&(\partial^{\mu} {\bar C}^{\nu} - \partial^{\nu}{\bar C}^{\mu}) \partial_{\nu}\beta -(\partial^{\mu} C^{\nu} - \partial^{\nu} C^{\mu}) B_{\nu} -\lambda B^{\mu}\nonumber\\
&-&\rho\partial^{\mu}\beta - \varepsilon^{\mu\nu\eta\kappa}(\partial_{\nu} C_{\eta})
{\cal B}_{\kappa},
\end{eqnarray}
\begin{eqnarray}
J^{\mu}_{(ab)}&=&\rho {\bar B}^{\mu} -(\partial^{\mu} {\bar C}^{\nu} - \partial^{\nu}{\bar C}^{\mu}) {\bar B}_{\nu} -(\partial^{\mu} C^{\nu} - \partial^{\nu} C^{\mu}) \partial_{\nu}{\bar\beta} \nonumber\\
&-& \lambda \partial^{\mu}{\bar\beta}- \varepsilon^{\mu\nu\eta\kappa}(\partial_{\nu}
{\bar C}_{\eta}){\bar{\cal B}}_{\kappa},
\end{eqnarray}
are derived from the Lagrangian densities (13) and (14), respectively.
Their conservation law can be proven by using the following equations of motion
\begin{eqnarray}
&& \lambda =\frac{1}{2}(\partial \cdot C), \quad  
\rho = -\frac{1}{2} (\partial \cdot {\bar C}),
\quad \Box \beta = 0, \quad \Box\bar\beta = 0, \quad 
\Box \lambda = 0,\nonumber\\
&& \Box {\bar C}_{\mu} =\frac{1}{2} \partial_{\mu}(\partial \cdot \bar C)
\equiv -\partial_{\mu}\rho, \quad
 \Box C_{\mu} =\frac{1}{2} \partial_{\mu}(\partial \cdot C)\equiv \partial_{\mu}\lambda, \quad\Box \rho =0, \nonumber\\
&& \varepsilon_{\mu\nu\eta\kappa}\partial^{\eta} {\cal B}^{\kappa} + (\partial_{\mu} B_{\nu} -\partial_{\nu} B_{\mu}) \equiv \varepsilon_{\mu\nu\eta\kappa}\partial^{\eta} B^{\kappa} - 
(\partial_{\mu} {\cal B}_{\nu} - \partial_{\nu} {\cal B}_{\mu}) = 0,\nonumber\\
&& \varepsilon_{\mu\nu\eta\kappa} \partial^{\eta} {\bar{\cal B}}^{\kappa} 
+ (\partial_{\mu} \bar B_{\nu} - \partial_{\nu} {\bar B}_{\mu}) \equiv 
\varepsilon_{\mu\nu\eta\kappa} \partial^{\eta} {\bar B}^{\kappa} - 
(\partial_{\mu} {\bar{\cal B}}_{\nu} - \partial_{\nu} {\bar{\cal B}}_{\mu}) = 0,
\end{eqnarray}
in addition to the equations enumerated in (8), (11) and (12).

The expression for the conserved (anti-)BRST charges are as follows:
\begin{eqnarray}
Q_{ab} &=& {\displaystyle \int}\; d^{3} x \Bigr[\rho {\bar B}^{0} -(\partial^{0} {\bar C}^{i} - \partial^{i}
{\bar C}^{0}) {\bar B}_{i} -(\partial^{0} C^{i} - \partial^{i} C^{0}) \partial_{i}{\bar\beta} \nonumber\\
&-& \lambda \partial^{0}{\bar{\beta}}-\varepsilon^{ijk}(\partial_{i} {\bar C}_{j}){\bar{\cal B}}_{k} \Bigl ],
\end{eqnarray} 
\begin{eqnarray}
Q_b &=& {\displaystyle \int}\;
d^{3} x \Bigr [(\partial^{0} {\bar C}^{i} - \partial^{i}{\bar C}^{0}) \partial_{i}\beta -(\partial^{0} C^{i} - \partial^{i} C^{0}) B_{i} -\lambda B^{0}\nonumber\\
&-&\rho\partial^{0}\beta - \varepsilon^{ijk}(\partial_{i} C_{j}){\cal B}_{k} \Bigl ].
\end{eqnarray} 
The above charges generate the (anti-)BRST transformations (16) and (15) as can be checked from the following relationship 
\begin{eqnarray}
s_r \Phi = -i \; \bigl [\Phi, Q_r \bigr ]_{[(\pm)]}, \qquad r = b, ab,
\end{eqnarray} 
where the [(+)-] signs, as the subscript on the square bracket, correspond to the (anti)commutator for the generic field $\Phi$
of the Lagrangian densities (13) and (14) being (fermionic)bosonic in nature.

The following algebraic structure
\begin{eqnarray}
s_{b} Q_{b} &=& -i \{ Q_{b}, Q_{b} \} = 0, \qquad s_{ab} Q_{ab} = -i \{ Q_{ab}, Q_{ab} \} = 0,\nonumber\\
s_{b} Q_{ab} &=& -i \{ Q_{ab}, Q_{b} \} = 0, \qquad s_{ab} Q_{b} = -i \{ Q_{b}, Q_{ab} \} = 0,
\end{eqnarray}
is derived from the transformations (15) and (16) when we exploit the expressions $Q_{(a)b}$ 
from (25) and (26) and use the definition of the generator from (27). The proof of 
$\{ Q_b, Q_{ab}\} = 0$, from the above equation (28), is a bit more involved. Some of 
the key algebraic steps are given in our Appendix A.\\

\noindent
{\bf 4 Absolutely anticommuting (anti-)co-BRST transformations}\\

\noindent
In this section, we shall show that, in addition to the transformations (15) and (16), there are other fermionic type {\it symmetry} transformations for the Lagrangian densities (13) and (14), under which, the total {\it gauge-fixing}  term remains invariant. These 
(anti-)co-BRST symmetry transformations (i.e. $s_{(a)d}$)  
are off-shell nilpotent ($ s_{(a)d}^2 = 0$) as can seen from the following:
\begin{eqnarray}
&& s_d B_{\mu\nu} = -\varepsilon_{\mu\nu\eta\kappa} \partial^\eta  \bar C^\kappa, \qquad s_d \bar C_\mu = - \partial_\mu \bar\beta, \qquad s_d C_\mu = - {\cal B}_\mu, \nonumber\\
&& s_d \phi_{2} = 2\rho, \quad s_d  \beta = - \lambda, \quad s_d \bigl [\rho, \lambda, 
\bar \beta, \phi_{1}, {\cal B}_{\mu}, B_\mu, \partial^\nu B_{\nu\mu} \bigr ] = 0,
\end{eqnarray}
\begin{eqnarray}
&& s_{ad} B_{\mu\nu} = -\varepsilon_{\mu\nu\eta\kappa} \partial^\eta  C^\kappa, \qquad s_{ad}  C_\mu =  \partial_\mu \beta, \qquad s_{ad} \bar C_\mu =  \bar{\cal B}_\mu, \nonumber\\
&& s_{ad} \phi_{2} = 2\lambda, \quad s_{ad} \bar\beta = \rho,  
\quad s_{ad} \bigl [\rho, \lambda, \beta, \phi_{1}, \bar{\cal B}_{\mu}, 
\bar B_\mu, \partial^\nu B_{\nu\mu} \bigr ] = 0.
\end{eqnarray}
The above transformations are the {\it symmetry} transformations as is evident from the following 
forms of the change in the Lagrangian densities:
\begin{eqnarray}
s_{d}{\cal L}_{(B,{\cal B})} = \partial_{\mu} \Bigr [(\partial^\mu  \bar C^\nu - \partial^\nu \bar C^\mu) {\cal B}_{\nu} - \lambda \;\partial^{\mu} \bar \beta - \rho {\cal B}^{\mu} \Bigl ], \nonumber\\
s_{ad}{\cal L}_{(\bar B,{\bar{\cal B}})} = \partial_{\mu} \Bigr [(\partial^\mu C^\nu - \partial^\nu C^\mu) \bar{\cal B}_{\nu} + \rho \; \partial^{\mu} \beta  + \lambda \bar{\cal B}^ \mu \Bigl ].
\end{eqnarray}
The above expressions show that the Lagrangian densities (13) and (14) are quasi-invariant under the transformations (29) and (30).

The Noether's conserved currents, corresponding to the nilpotent
and continuous symmetry transformations (29) and (30), are as follows:
\begin{eqnarray}
J^{\mu}_{(d)}&=&(\partial^{\mu} {\bar C}^{\nu} - \partial^{\nu}{\bar C}^{\mu}) {\cal B}_\nu 
- (\partial^{\mu} C^{\nu} - \partial^{\nu} C^{\mu})\; \partial_\nu  \bar \beta 
- \rho {\cal B}^{\mu}\nonumber\\
&-& \lambda \partial^{\mu}\bar\beta 
- \varepsilon^{\mu\nu\eta\kappa}(\partial_{\nu} \bar C_\eta) B_{\kappa},
\end{eqnarray}
\begin{eqnarray}
J^{\mu}_{(ad)}&=&(\partial^{\mu} C^{\nu} - \partial^{\nu} C^{\mu}) \bar {\cal B}_\nu - (\partial^{\mu} \bar C^{\nu} - \partial^{\nu} \bar C^{\mu}) \partial _\nu \beta 
+ \lambda \bar {\cal B}^{\mu}\nonumber\\
&+& \rho  \partial^{\mu}\beta - 
\varepsilon^{\mu\nu\eta\kappa}(\partial_{\nu} C_\eta) \bar B_{\kappa}.
\end{eqnarray}
The conservation law (i.e. $\partial_\mu J^\mu_{(a)d} = 0$) can be proven by exploiting the equations of motion (8), (11), (12) and (24).

The generators of the above nilpotent transformations (29) and (30) can be calculated from the above conserved currents as 
\begin{eqnarray}
Q_d = {\displaystyle \int}\; d^3 x J^0_{(d)} &=& {\displaystyle \int}\;
d^{3} x \Bigr [(\partial^{0} {\bar C}^{i} - \partial^{i}{\bar C}^{0}) {\cal B}_i -(\partial^{0} C^{i} - \partial^{i} C^{0}) \partial_i \bar\beta  \nonumber\\
&-&\rho {\cal B}^{0} -\lambda \partial^{0}\bar \beta - \varepsilon^{ijk}(\partial_{i} \bar C_{j}) B_{k} \Bigl ],
\end{eqnarray}
\begin{eqnarray}
Q_{ad} = {\displaystyle \int}\; d^3 x J^0_{(ad)} &=& {\displaystyle \int}\; 
d^{3} x \Bigr [(\partial^{0} C^{i} - \partial^{i} C^{0}) \bar {\cal B}_i -(\partial^{0} \bar C^{i} - \partial^{i} \bar C^{0}) \partial_i \beta  \nonumber\\
&+& \lambda \bar{\cal B}^{0} + \rho \partial^{0} \beta - \varepsilon^{ijk}(\partial_{i} C_{j}) \bar B_{k} \Bigl ].
\end{eqnarray}
We christen the above conserved charges as the co-BRST and anti-co-BRST.

Exploiting the canonical brackets, associated with the Lagrangian densities (13) and (14), it can be shown that $Q_{(a)d}^2 = 0$ and $\{ Q_d,Q_{ad}\} = 0$. However, it can be checked that
the following relationships are true, namely;
\begin{eqnarray}
&& s_d Q_d = - i \{ Q_d, Q_d \} = 0, \qquad  s_{ad} Q_{ad} = - i \{ Q_{ad}, Q_{ad} \} = 0,
\nonumber\\
&& s_d Q_{ad} = - i \{ Q_{ad}, Q_d \} = 0,  \qquad s_{ad} Q_d = - i \{ Q_d, Q_{ad} \} = 0,
\end{eqnarray}
due to the identification of (35) and (34) as the generators of the nilpotent (anti-)dual-BRST
symmetry transformations (30) and (29).

Before we close this section, a few side remarks are in order. First, 
the anticommutativity of the transformations $s_{(a)d}$ on the gauge field
\begin{eqnarray}
\{ s_d,  s_{ad} \} \; B_{\mu\nu} = \varepsilon_{\mu\nu\eta\kappa} \partial^\eta 
({\cal B}^\kappa - \bar {\cal B}^\kappa) = 0,
\end{eqnarray}
is satisfied  on the constrained surface defined by the field equation 
${\cal B}_\mu - \bar {\cal B}_\mu = \partial_\mu \phi_2$ given in (12). Second, it can be checked explicitly that the transformations
(29) and (30) are anticommuting on the rest of the fields of the theory if we include the 
following transformations on the auxiliary fields\footnote{Using the 
nilpotent transformations (29), (30) and
(38), it can be readily verified that
$s_{ad} {\cal L}_{(B, {\cal B})} 
=  \partial_\mu [ (\partial^\mu  C^\nu - \partial^\nu  C^\mu) {\cal B}_\nu 
+ \rho \partial^\mu  \beta  + \lambda (\bar {\cal B}^\mu + \frac{1}{2} 
\varepsilon^{\mu\nu\eta\kappa} \partial_\nu B_{\eta\kappa}) ]
- (\partial^\mu  C^\nu - \partial^\nu C^\mu) \partial_\mu ({\cal B}_\nu - \bar {\cal B}_\nu)
- (\partial^\mu \lambda) (\bar {\cal B}_\mu + 2 {\cal B}_\mu - \frac{1}{2} \partial_\mu \phi_2)$ 
and
$s_{d} {\cal L}_{(\bar B, \bar {\cal B})} 
=  \partial_\mu [ (\partial^\mu  \bar C^\nu - \partial^\nu \bar C^\mu) \bar {\cal B}_\nu 
- \rho ( {\cal B}^\mu + \frac{1}{2} \varepsilon^{\mu\nu\eta\kappa}
\partial_\nu B_{\eta\kappa})  - \lambda \partial^\mu \bar \beta ]
+ (\partial^\mu \bar C^\nu - \partial^\nu \bar C^\mu) \partial_\mu 
({\cal B}_\nu - \bar {\cal B}_\nu)
+ (\partial^\mu \rho) ( {\cal B}_\mu + 2 \bar {\cal B}_\mu + \frac{1}{2} \partial_\mu \phi_2)$.
Due to the relations (12), it can be verified that
the Lagrangian densities (13) and (14) remain invariant under the
transformations (30) and (29), respectively, on the constrained surface where
${\cal B}_\mu - \bar {\cal B}_\mu = \partial_\mu \phi_2$, $\bar {\cal B}_\mu + 2 {\cal B}_\mu - \frac{1}{2} \partial_\mu \phi_2 = \frac{3}{2} \varepsilon_{\mu\nu\eta\kappa}
\partial^\nu B^{\eta\kappa}, 
{\cal B}_\mu + 2 \bar {\cal B}_\mu + \frac{1}{2} \partial_\mu \phi_2 = \frac{3}{2} \varepsilon_{\mu\nu\eta\kappa}
\partial^\nu B^{\eta\kappa} 
$.}
\begin{eqnarray}
s_d \bar {\cal B}_\mu = \partial_\mu \rho, \quad  s_{ad} {\cal B}_\mu = - \partial_\mu \lambda, 
\quad s_{ad}  B_\mu = 0, \quad s_d \bar B_\mu = 0, 
\end{eqnarray}
in addition to (29) and (30). Finally, in the proof of $s_d Q_{ad} = - i \{Q_{ad}, Q_d \} = 0$
as well as $s_{ad} Q_d = - i \{Q_d, Q_{ad} \} = 0$, one has to exploit both the constrained
field equations $B_\mu - \bar B_\mu = \partial_\mu \phi_1$ as well as 
${\cal B}_\mu - \bar {\cal B}_\mu = \partial_\mu \phi_2$ when we compute the explicit
expressions $s_d Q_{ad}$ and $s_{ad} Q_d$ by using (29), (30), (34) and (35). One such
computation, in a concise manner, is illustrated in our Appendix B.\\

\noindent
{\bf 5 Bosonic symmetries: analogue of the Laplacian operator}\\

\noindent
It is clear, from the previous sections, that we have four nilpotent symmetry transformations
(i.e. $s_{(a)b}, s_{(a)d}$) in our present theory. We have also shown that the following anticommutators are true, namely;
\begin{eqnarray}
\{ s_b, s_{ab} \} = 0, \quad \{ s_d, s_{ad} \} = 0, \quad 
\{ s_b, s_{ad} \} = 0, \quad \{ s_d, s_{ab} \} = 0,
\end{eqnarray}
on the constrained surface defined by the field equations in (12).

At this juncture, it is very natural to expect that the remaining 
non-zero pair of the anticommutators,
as listed below: 
\begin{eqnarray}
s_\omega = \{ s_b, s_d \}, \qquad  s_{\bar\omega} = \{ s_{ab}, s_{ad} \},
\end{eqnarray}
would correspond to the bosonic 
symmetry transformations in the theory because $s_{a(d)}$ and $s_{a(b)}$ are individually nilpotent
(fermionic) symmetry transformations. A close look at (17) and (31) certify the above assertions.

The following bosonic transformations $s_\omega = \{ s_b, s_d \}$: 
\begin{eqnarray}
&& s_\omega B_{\mu\nu} = \partial_\mu {\cal B}_\nu -\partial_\nu {\cal B}_\mu + \varepsilon_{\mu\nu\eta\kappa} \partial^\eta B^\kappa , \quad s_\omega C_\mu = \partial_\mu \lambda, \nonumber\\
&& s_\omega \bar C_\mu = \partial_\mu \rho, \quad s_\omega \Bigl [ \phi_1, \phi_2,
\beta, \bar\beta, \lambda, \rho, B_\mu, \bar B_\mu, {\cal B}_\mu, 
{\bar {\cal B}_\mu} \Bigr ] = 0,
\end{eqnarray}
are the symmetry transformations of the Lagrangian density (13) because:
\begin{eqnarray}
s_\omega {\cal L}_{(B,{\cal B})} &=& \partial_\mu \Bigl [ {\cal B}^\mu (\partial \cdot B)
- B^\mu (\partial \cdot {\cal B}) + B^\nu \partial^\mu {\cal B}_\nu
- {\cal B}^\nu \partial^\mu  B_\nu \nonumber\\
&+& (\partial ^\mu \lambda ) \rho - \lambda (\partial^\mu \rho ) \Bigr ],
\end{eqnarray}
shows that the Lagrangian density
$ {\cal L}_{(B, {\cal B})}$ remains quasi-invariant under (41).

Exactly, in the above manner, it can be checked that following bosonic 
infinitesimal transformations $s_{\bar \omega} = \{ s_{ab}, s_{ad} \}$: 
\begin{eqnarray}
&& s_{\bar\omega} B_{\mu\nu} = - (\partial_\mu \bar {\cal B}_\nu -\partial_\nu \bar {\cal B}_\mu + \varepsilon_{\mu\nu\eta\kappa} \partial^\eta \bar B^\kappa), \quad s_{\bar\omega} C_\mu = 
- \partial_\mu \lambda, \nonumber\\
&& s_{\bar\omega} \bar C_\mu = - \partial_\mu \rho, \quad s_{\bar\omega} \Bigl [ \phi_1, \phi_2,
\beta, \bar\beta, \lambda, \rho, B_\mu, \bar B_\mu, {\cal B}_\mu, 
{\bar {\cal B}_\mu} \Bigr ] = 0,
\end{eqnarray}
leave the Lagrangian density (14) (i.e.  $ {\cal L}_{(\bar B, \bar{\cal B})}$)
quasi-invariant as the latter transforms in the following fashion:
\begin{eqnarray}
s_{\bar\omega} {\cal L}_{(\bar B,\bar {\cal B})} &=& \partial_\mu 
\Bigl [ \bar  B^\mu (\partial \cdot \bar {\cal B})
- \bar {\cal B}^\mu (\partial \cdot \bar B)  + \bar {\cal B}^\nu \partial^\mu 
 \bar B_\nu  - \bar B^\nu \partial^\mu \bar {\cal B}_\nu   \nonumber\\
&+& \lambda (\partial^\mu \rho) - (\partial^\mu \lambda ) \rho \Bigr ].
\end{eqnarray}
Thus, equations (42) and (44) imply that $ s_\omega $ and $ s_{\bar \omega} $
are the {\it symmetry} transformations for the Lagrangian densities (13) and (14), 
respectively. These symmetry transformations owe their origin to the four basic fermionic 
(anti-)BRST and (anti-)co-BRST symmetry transformations of the theory.

On their face value, the transformations (41) and (43) look completely independent.
However, a close observation of (41) and (43), using the constrained  field equations
(12), reveal that they differ only by a sign factor. To be specific, using
$B_\mu - \bar B_\mu = \partial_\mu \phi_1, {\cal B}_\mu - \bar {\cal B}_\mu = \partial_\mu
\phi_2$, it can be seen that $s_\omega + s_{\bar \omega} = 0$. Thus, there is nothing 
profound in the observation that $ [ s_\omega, s_{\bar \omega} ] \Phi = 0$ for the 
generic field $\Phi$ of the Lagrangian densities (13) and (14). It is a sheer coincidence that
in (41) and (43), we observe that $s_{\omega}^2 = 0$ and $s_{\bar\omega}^2 = 0$. Strictly
speaking, however, these transformations are bosonic.

The following Noether conserved current emerges when we exploit the 
continuous symmetry transformations (41):
\begin{eqnarray}
J^\mu_{(\omega)} &=& \varepsilon^{\mu\nu\eta\kappa}
\bigl \{ (\partial_\nu {\cal B}_\eta) {\cal B}_\kappa + (\partial_\nu B_\eta) B_\kappa
\bigr  \} + \partial_\nu  \bigl [ B^\mu {\cal B}^\nu - {\cal B}^\mu B^\nu \bigr ] \nonumber\\
&+& (\partial^\mu \bar C^\nu - \partial^\nu \bar C^\mu) \partial_\nu \lambda
- (\partial^\mu C^\nu - \partial^\nu  C^\mu) \partial_\nu \rho.
\end{eqnarray}
The conservation law (i.e. $\partial_\mu J^\mu_{(\omega)} = 0$) of the above current
can be proven by exploiting the equations of motion (8), (11), (12) and (24).

The conserved charge, corresponding to the above conserved current, is
\begin{eqnarray}
W = {\displaystyle \int} d^3 x  J^0_{(\omega)} &\equiv& {\displaystyle \int} d^3 x 
\Bigl [ \epsilon^{ijk} \bigl \{ (\partial_i {\cal B}_j) {\cal B}_k +
 (\partial_i  B_j) B_k \bigr \} 
+ (\partial^0 \bar C^i - \partial^i \bar C^0) \partial_i \lambda  \nonumber\\
&-& (\partial^0  C^i - \partial^i  C^0) \partial_i \rho \Bigr ].
 \end{eqnarray}
There are other ways to compute this conserved  charge. For instance, it can be
checked that $s_b Q_d = - i \{Q_d, Q_b \}, s_d Q_b = - i \{Q_b, Q_d \}$ 
can be used to deduce
the expression for $W$. Similarly the expressions
$s_{ab} Q_{ad} = - i \{Q_{ad}, Q_{ab} \}$ and $
 s_{ad} Q_{ab} = - i \{Q_{ab}, Q_{ad} \}$
lead to the derivation of $W$.
Some subtlety of these computations are discussed briefly in our Appendix C.\\

\noindent
{\bf 6 Ghost and discrete symmetry transformations: the ghost charge}\\

\noindent
It will be noted that the ghost part of the Lagrangian densities (13) and (14)
\begin{eqnarray}
{\cal L}_{(g)} = \partial_{\mu}{\bar \beta } \partial^{\mu} \beta 
+ (\partial_{\mu} {\bar C}_{\nu} - \partial_{\nu}{\bar C}_{\mu})(\partial^{\mu}C^{\nu})
+(\partial \cdot C - \lambda)\rho + (\partial \cdot {\bar C}+ \rho )\lambda,
\end{eqnarray}
respects the following continuous global ($\Sigma \neq \Sigma (x)$) scale symmetry
transformations for the ghost fields of the theory, namely;
\begin{eqnarray}
&& C_\mu \rightarrow e^{+ \Sigma} C_\mu, \quad \bar C_\mu \rightarrow e^{- \Sigma} \bar C_\mu, \quad \beta \rightarrow e^{+2 \Sigma} \beta, \nonumber\\
&& \bar \beta \rightarrow e^{-2 \Sigma} \bar \beta, \quad \rho \rightarrow e^{- \Sigma} \rho, \quad \lambda \rightarrow e^{+ \Sigma} \lambda,
\end{eqnarray}
where numbers $ (\pm 1)$ and $(\pm 2)$, in the exponentials, stand for the ghost numbers
of the corresponding (anti-)ghost fields. It is evident that $ \lambda $ and $ \rho $ 
have the ghost number (+1) and (-1), respectively, because of the fact that  
$ \lambda = + \frac{1}{2}(\partial \cdot C), \rho = -\frac {1}{2} (\partial \cdot \bar C)$.
Furthermore, the ghost number for the rest of the fields of the theory 
(i.e. $B_{\mu\nu}, B_\mu, \bar B_\mu ,{\cal B}_\mu, \bar {\cal B}_\mu ,\phi_1, \phi_2)$
is zero. Thus, under ghost symmetry transformations :
$ B_{\mu\nu} \rightarrow B_{\mu\nu}, B_\mu \rightarrow  B_\mu, \bar B_\mu \rightarrow \bar B_\mu  ,{\cal B}_\mu \rightarrow {\cal B}_\mu, \bar {\cal B}_\mu \rightarrow \bar {\cal B}_\mu,\phi_1 \rightarrow \phi_1 $ and  $\phi_2 \rightarrow \phi_2$.

The infinitesimal version  (i.e. $\Sigma \rightarrow 0)$ of the above 
global scale transformations 
(i.e. $s_g)$ is as given below:
\begin{eqnarray}
&& s_g C_\mu = + \Sigma C_\mu,\quad  s_g \bar C_\mu = - \Sigma \bar C_\mu, \quad 
s_g \rho = - \Sigma \rho, \nonumber\\
&& s_g \lambda = + \Sigma \lambda, \quad s_g \beta = + 2 \Sigma \beta, \quad
s_g \bar \beta = - 2 \Sigma \bar \beta.
\end{eqnarray}
The above symmetry transformations lead to the derivation of the 
conserved Noether current (i.e. the ghost current) as:
\begin{eqnarray}
J^\mu_{(g)} &=& 2 \beta \partial^\mu \bar \beta - 2 \bar \beta \partial^\mu \beta
+ ( \partial^\mu C^\nu -\partial^\nu C^\mu )\bar C_\nu \nonumber\\
&+& ( \partial^\mu \bar C^\nu -\partial^\nu \bar C^\mu ) C_\nu + C^\mu \rho
- \bar C^\mu \lambda .
\end{eqnarray}
The conservation law $ ( \partial_\mu J^\mu_{(g)} = 0 ) $ can be readily 
proven by exploiting the equations of motion for the (anti-)ghost fields from (24).

The generator of the infinitesimal transformations (49) is the conserved 
(i.e. $\dot Q_{(g)} = 0$) ghost charge $ Q_g$ defined by the following expression:
\begin{eqnarray}
Q_g = \int d^3 x  J^0_{(g)} &=& \int d^3 x \Bigl [ 2 \beta \partial^0 \bar \beta - 2 \bar \beta \partial^0 \beta + ( \partial^0 C^i -\partial^i C^0 )\bar C_i \nonumber\\
&+& ( \partial^0 \bar C^i -\partial^i \bar C^0 ) C_i + C^0 \rho
- \bar C^0 \lambda \Bigr ].
\end{eqnarray}
Exploiting the infinitesimal
transformations (15), (16), (29), (30), (41) and (49)
(with $\Sigma = 1$), the following algebraic structure
can be deduced:
\begin{eqnarray}
&& s^2_{(a)b} = 0, \; s^2_{(a)d} = 0, \; \{ s_b, s_{ab} \} = 0, \; \{ s_d, s_{ad} \} = 0,\
\{ s_d, s_b \} = s_\omega, \nonumber\\
&& \{ s_{ad},s_{ab} \} = s_{\bar\omega} \equiv - s_\omega, \quad  
[ s_\omega, s_r] = 0, \quad r = b, ab, d, ad, g,  \nonumber\\
&& [s_g, s_b] =+ s_b, \; [s_g, s_d] = - s_d, \; [s_g, s_{ab}] = - s_{ab}, 
\; [s_g, s_{ad}] = + s_{ad}.
\end{eqnarray}
All the rest of the (anti)commutators of the above infinitesimal
transformations (e.g. $[s_g, s_g] = 0$, etc.) are found to be trivially zero.

In addition to the continuous symmetry transformations (49), the ghost part of the 
Lagrangian density (i.e. the equation (47)) respects the following discrete 
symmetry transformations:
\begin{eqnarray}
&& C_\mu \rightarrow \pm i \bar C_\mu, \quad \bar C_\mu \rightarrow \pm i C_\mu,
\quad \beta \rightarrow \pm i \bar \beta, \nonumber\\
&& \bar \beta \rightarrow \mp i \beta, \quad 
\rho \rightarrow \mp i\lambda, \quad \lambda \rightarrow \mp i \rho.
\end{eqnarray}
Thus, we note that, under the discrete symmetry transformations (9), (10) and (53),
the total Lagrangian densities (13) and (14) remain invariant.

The discrete symmetry transformations (9), (10) and (53), combined together, 
correspond to the Hodge duality $ * $ operation of differential geometry. To 
corroborate this assertion, it is essential to note that a pair of above discrete 
transformations, on the bosonic (B) and fermionic (F) fields of the theory, 
lead to the following expressions [28]:
\begin{eqnarray}
&& * ( * B ) = + B, \qquad  B = B_{\mu\nu}, B_\mu, \bar B_\mu, {\cal B}_\mu, 
\bar{\cal B}_\mu, \phi_1, \phi_2, \beta, \bar\beta, \nonumber\\
&& * ( * F )  = - F, \qquad  F = C_\mu, \bar C_\mu, \rho, \lambda .
\end{eqnarray}
The above signs are important for our purpose because it can be seen that, 
in the following relationships\footnote{It can be checked
that the relation $s_{(a)b} \Phi = \mp * s_{(a)d} * \Phi$ (with 
$\Phi = B, F$) is also true. Here $ B$ and $F$ are defined in (54).
The sign-flip on the r.h.s. is due
to the dimensionality of the spacetime manifold on which the fields
of the theory are defined (see, e.g. [28]).} (see, e.g. [28] for details)
\begin{eqnarray}
s_{(a)d} \Phi = \pm * s_{(a)b} * \Phi, 
\qquad \Phi = B, F,                                                                      \end{eqnarray}
the (+)- signs are dictated by the signs in (54). The above relationship
is the analogue of the relationship $ \delta = \pm * d * $  that exists between the
exterior and co-exterior derivatives $ d $ and $\delta$ of differential geometry.

It is worthwhile to point out that, in the realm of differential geometry
on a compact manifold without a boundary, the signature in the relationship
$ \delta = \pm * d * $ is decided by the dimension of the manifold and the 
degree of the differential forms that are involved in the inner product.
For instance, for the even dimensional compact manifold $ \delta = - * d * $
is always true (see, e.g. [3,4] for details). In the realm of BRST 
formalism for the 4D free Abelian 2-form gauge theory, the signature in (55) 
is dictated by (54).

Before we close this section, it is interesting to note that the application of 
the discrete symmetry transformations (9), (10) and (53) (i.e. the analogue 
of the $ * $ operation) on the conserved charges is
\begin{eqnarray}
&& * \; Q_b = Q_d, \qquad * \; Q_d = - Q_b, \qquad * \; W = - W, \nonumber\\
&& * \; Q_{ad} = - Q_{ab}, \qquad * \; Q_{ab} =  Q_{ad}, \qquad * \;  Q_g = - Q_g.
\end{eqnarray}
The above equation shows that, under the discrete symmetry transformations 
of the theory, $ Q_b \rightarrow  Q_d, Q_d \rightarrow - Q_b $ (and
$ Q_{ab} \rightarrow Q_{ad}, Q_{ad} \rightarrow - Q_{ab} $) which are exactly like the
electromagnetic duality  (i.e. $\vec E \rightarrow \vec B, \vec B \rightarrow - \vec E $ )
that exists for the Maxwell's source free field equations. \\

\noindent
{\bf 7 Algebraic structures: cohomological aspects}\\ 

\noindent
The algebra obeyed by the transformations $ s_r $ (with $ r = b, ab, d, ad, \omega, g $)
is replicated by the generators of these transformations. Exploiting the canonical 
(anti)commutators, derived from the Lagrangian densities (13) and (14), it can be 
shown that the following algebraic structure is true, namely;
\begin{eqnarray}
&& Q^2_{(a)b} = 0, \quad Q^2_{(a)d} = 0,\quad  [W, Q_r] = 0, \quad ( r = b, ab, d, ad, 
g),\nonumber\\
&& \{ Q_b, Q_{ab} \} = 0, \quad \{ Q_d, Q_{ad} \} = 0,
\quad  \{ Q_b, Q_{ad} \} = 0,  \nonumber\\
&& \{ Q_d, Q_{b} \} = - \{ Q_{ad}, Q_{ab} \} = W,     
\quad \{ Q_d, Q_{ab} \} = 0,\nonumber\\
&& i [ Q_g, Q_b ] = + Q_b, \quad i [ Q_g, Q_{ab} ] = - Q_{ab}, \nonumber\\
&& i [ Q_g, Q_d ] = - Q_d, \quad i [ Q_g, Q_{ad} ] = + Q_{ad}.
\end{eqnarray}
This is the extended BRST algebra corresponding to our present 4D Abelian 2-form
gauge theory which is endowed with six symmetry transformations.

We can define the ghost number of a state (in the quantum Hilbert space of states) 
as the eigen value of the operator $ i Q_g $. In other words, a state $|\psi >_n $
(with $ i Q_g | \psi >_n  = n |\psi >_n  $ ) has the ghost number $n$.
As a result of the algebra in (57), it can be checked that the following
relations are true:
\begin{eqnarray}
&& i Q_g Q_b | \psi >_n = (n + 1) Q_b | \psi >_n, \quad 
i Q_g Q_d | \psi >_n = (n - 1) Q_d | \psi >_n, \nonumber\\
&& i Q_g Q_{ab} | \psi >_n = (n - 1) Q_{ab} | \psi >_n,\;
i Q_g Q_{ad} | \psi >_n = (n + 1) Q_{ad} | \psi >_n, \nonumber\\
&& i Q_g W | \psi >_n = n W | \psi >_n.
\end{eqnarray}
The above relationships demonstrate that the ghost numbers of the states $ Q_b |\psi >_n, 
Q_d | \psi > _n $
and $ W |\psi >_n $ are $(n + 1), (n - 1)$ and $n$ respectively. In exactly similar 
fashion, the states $ Q_{ad} |\psi >_n$ and $ Q_{ab} |\psi >_n $ have ghost numbers
$(n + 1)$ and $(n - 1)$, respectively.

The structure of the algebra in (57) and the relationship in  (58) demonstrate that
there are two sets of generators of transformations that correspond to the de Rham
cohomological differential operators $ d, \delta, \Delta $. For instance, the first
set $ ( Q_b, Q_d, W ) $ and the second set $ ( Q_{ad}, Q_{ab}, - W ) $ obey exactly 
the same kind of algebra as $ ( d, \delta, \Delta ) $ which is:
$d^2 = \delta^2 = 0, \Delta = \{d, \delta \} = (d + \delta)^2, [\Delta, d ] = 0, 
[\Delta, \delta ] = 0$. Thus, the mapping is two-to-one
from the conserved charges (corresponding to the symmetries 
of the 2-form theory) to the cohomological operators
(of differential geometry on the compact manifolds), namely;
$ ( Q_b, Q_{ad} ) \rightarrow d, ( Q_d, Q_{ab} ) \rightarrow \delta $ and 
$( + W, - W) \rightarrow \Delta $.

It is well-known that the exterior derivative raises the degree of a form by one
when it operates on it. On the other hand, the dual-exterior derivative lowers
the degree of a form by one due to its action on the latter. These properties 
of $d$ and $\delta$ are mimicked by sets $(Q_b, Q_{ad})$ and $(Q_d, Q_{ab})$,
respectively. As is evident from (58), the set $(Q_b, Q_{ad})$ raises the ghost
number of a state by one and the set $(Q_d, Q_{ab})$ lowers the ghost number 
of the same state by one. Furthermore, it is an important point to note that $Q_b$ and
$Q_{ab}$ are independent of each-other (i.e. $\{Q_b, Q_{ab} \} = 0$) as are
$Q_d$ and $Q_{ad}$ because of $\{ Q_d, Q_{ad} \} = 0$.
These observations enable us to express any arbitrary state $|\psi>_n$, due to the
Hodge decomposition theorem (HDT)\footnote{On a compact manifold without a boundary,
any arbitrary $n$-form $f_n$ can be uniquely written as the sum of the harmonic form
$h_n$ (with $\Delta h_n = 0, d h_n = 0, \delta h_n = 0$), an exact form
$(d e_{n-1})$ and a co-exact form $(\delta c_{n + 1})$. Thus, the HDT can
be mathematically expressed as: $f_n = h_n + d e_{n-1} + \delta c_{n + 1}$ where
$h_n$ is annihilated by $d$ and $\delta$ together.} [3-6], as follows
\begin{eqnarray}
|\psi>_n &=& |\omega>_{(n)} + Q_b\; |\chi>_{(n - 1)} + Q_d \;|\theta>_{(n + 1)} \nonumber\\
&\equiv&  |\omega>_{(n)} + Q_{ad} \;|\chi>_{(n - 1)} + Q_{ab} \;|\theta>_{(n + 1)},
\end{eqnarray}
where $|\omega>_n$ is the harmonic state, $Q_b |\chi>_{(n - 1)}$ is the BRST exact state
and $Q_d |\theta>_{(n + 1)}$ is the co-BRST exact state. In a similar fashion, the second
line of the above equation can also be defined.

In the above,
the most symmetric state is the harmonic state because it is (anti-)BRST as well as
(anti-)co-BRST invariant. This is why, it is appropriate to choose this state as 
the physical state of the theory. The physicality criteria (i. e. $Q_{(a)b} |phys> = 0,
Q_{(a)d} |phys> = 0$) on the physical state $|phys>$ of the theory leads to the annihilation
of the physical state by the operator form of the first-class constraints and their dual.
This analysis has already been performed in our earlier works [15]. Thus, we shall not
dwell on it in our present endeavour because the results are almost the same. \\

\noindent
{\bf 8 Conclusions}\\

\noindent
In our present investigation, we have demonstrated that the free 4D Abelian 2-form
gauge theory is a tractable field theoretical model for the Hodge theory because all
the de Rham cohomological operators of differential geometry 
find their physical realizations in the language of the
well-defined symmetry transformations of the specific Lagrangian densities (cf. (13) and (14))
of the theory. It turns out that the total kinetic term of the gauge field, owing its origin to
the exterior derivative $d = dx^\mu \partial_\mu$, remains invariant under the (anti-)BRST
symmetry transformations. On the other hand, the total gauge-fixing term, owing its
origin to the co-exterior derivative $\delta = \pm * d *$, is found to remain invariant
under the (anti-)co-BRST symmetry transformations\footnote{Besides the role of $d$ and $\delta$, there are other specific subtleties that are also 
involved in our discussion of the (anti-)BRST and (anti-)co-BRST 
transformations (cf. Sec. 2). For instance, the massless scalar fields $\phi_2$ and $\phi_1$
also remain invariant under the (anti-)BRST and (anti-)co-BRST symmetry transformations, 
respectively.}.

The discrete symmetry transformations (9), (10) and (53), present in our 
Abelian 2-form gauge theory, are found to be the realization of the
Hodge duality $*$ operation of the differential geometry in the relationship $\delta = \pm
* d *$. The interplay of the continuous symmetries and the discrete symmetries of the theory
encode the above relationship in an explicit manner as is evident from our equation (55). The
$(\pm)$ signs of the above relationship are captured in (54) where the operation of the
two successive discrete symmetry transformations on the fields 
of the theory, unambiguously decides it [28].

According to the Noether's theorem, the continuous symmetry transformations lead to the
conserved charges. We have six continuous symmetries in the theory which lead to six
conserved charges as $Q_b, Q_{ab}, Q_d, Q_{ad}, W, Q_g$. These charges
obey the algebra (57) that is reminiscent of the algebra
respected by the de Rham cohomological operators of  differential geometry. It turns out
that, under the duality transformations (9), (10) and (53), the algebraic structure in (57)
remains intact as is clear from the transformations (56). Thus, we conclude that the whole
0theory is duality invariant because (i) the Lagrangian densities (13) and
(14) of the theory remain invariant under (9), (10) and (53), and (ii) the algebraic structure
(57) {\it also} remains invariant under the discrete symmetry transformations (9), (10) and (53).

There are significant physical implications of our present kind of studies. For instance,
we have been able to demonstrate, because of the above type of studies,
that the two (1 + 1)-dimensional (2D) free Abelian and non-Abelian
gauge theories (having no interaction with matter fields) present a new type of topological
field theories which capture a part of the salient features of the Witten-type of
topological theories and some of the key
properties of the 
Schwarz-type of topological theories  (see, e.g. [13] for details). Furthermore, the
2D interacting Abelian U(1) gauge theory (i.e. QED) presents a field theoretical model
for the Hodge theory where the topological gauge field $A_\mu$ couples with the Noether conserved
current constructed with the help of Dirac fields [14]. In addition, such studies have
established that the free Abelian 2-form gauge theory is a quasi-topological field theory [16].

We have established, in our very recent work [29], that the simple 2D free Abelian U(1) gauge theory
is a  field theoretical model for the Hodge theory. In this work, we have demonstrated
the {\it usefulness} of the ordinary as well as the super de Rham cohomological operators where the latter cohomological operators are defined
on the (2, 2)-dimensional supermanifold. We have exploited the importance of the super
exterior derivative in deriving the nilpotent and absolutely anticommuting
(anti-)BRST symmetry transformations
and Curci-Ferrari type restriction for the 4D free Abelian gauge theory in [18].
It would be interesting  venture to tap the potential of the super co-exterior derivative
and super Laplacian operator, defined on the (4, 2)-dimensional supermanifold,
for the 4D Abelian 2-form gauge theory.

It would be a challenging endeavour to capture the main features of our present
investigation in the language of the Hamiltonian formalism 
where the constraint structure of the theory is emphasized [30-32]. The study of the topological
features of the Abelian 2-form gauge theory, with the help of absolutely anticommuting
(anti-)BRST as well as (anti-)co-BRST symmetry transformation, is yet another direction
for further investigation. The generalization  of our present results 
to the case of the 4D non-Abelian 2-form gauge theory is a demanding problem for
us. There are some interesting field theoretical models where the 2-form gauge potential
appears in a compelling manner [33,34]. It would be nice to study them within the
framework of the BRST formalism and look for the existence of 
dual-BRST type symmetry transformations.
All the above issues are being investigated at the moment and our results
would be reported in our future publications [35].\\

\noindent
{\bf Acknowledgement:}
Financial support from the
Department of Science and Technology (DST), Government of India,  
under the SERC project sanction grant No: - SR/S2/HEP-23/2006, is gratefully acknowledged.

\begin{center}
{\bf Appendix A}
\end{center}

\noindent
Here we furnish some of the key steps in proving the fact that $\{ Q_b, Q_{ab} \} = 0$
by exploiting the transformations (15) and the expression for $Q_{ab}$ from (25)
in the computation $s_b Q_{ab} = - i \{ Q_{ab}, Q_{b} \}$. It can be checked that
\begin{eqnarray}
&& s_b Q_{ab} = {\displaystyle \int} d^3 x \Bigl [ \epsilon^{ijk} (\partial_i B_j) \bar {\cal B}_k
+ (\partial^0 B^i - \partial^i B^0) \bar B_i\nonumber\\
&& + \rho \dot \lambda - \lambda \dot \rho - (\partial^0 \bar C^i - \partial^i \bar C^0)
\partial_i \lambda - (\partial^0  C^i - \partial^i  C^0)
\partial_i \rho \Bigr ]. 
\end{eqnarray}
Using the constraint field equation ${\cal B}_\mu - \bar {\cal B}_\mu = \partial_\mu \phi_2$
from (12) and equation of motion $\varepsilon^{\mu\nu\eta\kappa} \partial_\eta {\cal B}_\kappa
+ (\partial^\mu B^\nu - \partial^\nu B^\mu) = 0$ (which implies
$(\partial^0 B^i - \partial^i B^0) =  - \epsilon^{ijk} \partial_j {\cal B}_k$) from (24), 
the first two  terms of the above equation lead to 
\begin{eqnarray}
 {\displaystyle \int} d^3 x  \Bigl [ \epsilon^{ijk} \partial_i {\cal B}_j (B_k - \bar B_k) \Bigr ]
\equiv 
 {\displaystyle \int} d^3 x  \Bigl [ \epsilon^{ijk} \partial_i {\cal B}_j 
(B_k - \bar B_k - \partial_k \phi_1) \Bigr ].
\end{eqnarray}
This expression is zero on constrained surface defined
by the field equation $B_\mu - \bar B_\mu = \partial_\mu \phi_1$. The rest of the terms of (60)
are as follows
\begin{eqnarray}
{\displaystyle \int} d^3 x \Bigl [  \rho \dot \lambda - \lambda \dot \rho 
+ (\partial_0 \bar C_i - \partial_i \bar C_0)
\partial_i \lambda + (\partial_0  C_i - \partial_i  C_0)
\partial_i \rho \Bigr ]. 
\end{eqnarray}
Performing a partial integration and throwing away the total space derivative terms, the
above equation can be recast into the following form:
\begin{eqnarray}
{\displaystyle \int} d^3 x \Bigl [  \rho \dot \lambda - \lambda \dot \rho 
- (\partial_0 \partial_i \bar C_i - \partial_i \partial_i \bar C_0)
\lambda - (\partial_0 \partial_i  C_i - \partial_i \partial_i  C_0)
\rho \Bigr ]. 
\end{eqnarray}
The above equation, with the help of the equations of motion 
$\lambda = \frac{1}{2} (\partial \cdot C), \rho = - \frac{1}{2} (\partial \cdot \bar C)$ 
from (24), can be reduced to 
\begin{eqnarray}
{\displaystyle \int} d^3 x \Bigl [ \dot \lambda \rho - \dot \rho  \lambda 
- (\Box \bar C_0) \lambda - (\Box  C_0) \rho \Bigr ] = 0, 
\end{eqnarray}
where we have used  $\Box C_0 = \dot \lambda$ and
$\Box \bar C_0 = - \dot \rho$ from (24).

\begin{center}
{\bf Appendix B}
\end{center}

\noindent
We very concisely provide some key inputs for the proof of $\{Q_d, Q_{ad} \} = 0$
from the computation of $s_d Q_{ad} = - i \{Q_{ad}, Q_{d} \}$ by exploiting the transformations
(29) and expression for $Q_{ad}$ from (35). It can be seen that
\begin{eqnarray}
&& s_d Q_{ad} = {\displaystyle \int} d^3 x 
\Bigl [\epsilon^{ijk} (\partial_i {\cal B}_j) \bar B_k
- (\partial^0 {\cal B}^i - \partial^i {\cal B}^0) \bar {\cal B}_i \nonumber\\
&& + \rho \dot \lambda - \lambda
\dot \rho - (\partial^0 C^i - \partial^i C^0) \partial_i \rho - (\partial^0 \bar C^i - 
\partial^i \bar C^0) \partial_i \lambda \Bigr ]. 
\end{eqnarray}
The ghost part of the above expression can be easily shown to be equal to zero by
exploiting the equations of motion from (24) (e.g. $\Box C_\mu = \partial_\mu \lambda,
\Box \bar C_\mu = - \partial_\mu \rho$ and $\lambda = \frac{1}{2} (\partial \cdot C),
\rho = - \frac{1}{2} (\partial \cdot \bar C)$). Now the first two terms of (65) can be recast
into the following form
\begin{eqnarray}
{\displaystyle \int} d^3 x 
\Bigl [\epsilon^{ijk} (\partial_i {\cal B}_j) \bar B_k
- \epsilon^{ijk} (\partial_j B_k) \bar {\cal B}_i \Bigr ], 
\end{eqnarray}
where we have exploited an appropriate equation of motion from (24) which implies
that  
$(\partial^0 {\cal B}^i - \partial^i {\cal B}^0) = \epsilon^{ijk} \partial_j B_k$.
Performing a partial integration and using the constrained field equation $B_\mu - \bar B_\mu = \partial_\mu \phi_1$, the above expression can be put in the following form
\begin{eqnarray}
{\displaystyle \int} d^3 x 
\Bigl [\epsilon^{ijk} \;(\partial_i B_j)\; ({\cal B}_k - \bar {\cal B}_k) \Bigr ]
\equiv 
{\displaystyle \int} d^3 x 
\Bigl [\epsilon^{ijk} \;(\partial_i B_j)\; ({\cal B}_k - \bar {\cal B}_k 
- \partial_k \phi_2) \Bigr ],
\end{eqnarray}
which reduces to zero on the constrained  surface parametrized by the field
equation ${\cal B}_\mu - \bar {\cal B}_\mu = \partial_\mu \phi_2$ from (12).
Thus, we note that equations (12) and (24) play important roles in the proof
of $\{ Q_d, Q_{ad} \} = 0$.

\begin{center}
{\bf Appendix C}
\end{center}

\noindent
We give a synopsis of the other ways to compute the 
conserved bosonic charge $W$  of equation (46). Exploiting the BRST symmetry
transformation (15) and (21) and applying them
onto the expression of $Q_d$ from (34), it is clear that 
\begin{eqnarray}
s_b Q_d &=& {\displaystyle \int} d^3 x  \Bigl [\epsilon^{ijk} (\partial_i B_j) B_k
- \lambda \dot \rho - (\partial^0 C^i - \partial^i C^0) \partial_i \rho \nonumber\\
&-& (\partial^0 B^i - \partial^i B^0) {\cal B}_i \Bigr ].
\end{eqnarray}
Using the equations of motion from (24), it can be seen that
$ (\partial^0 B^i - \partial^i B^0) = - \epsilon^{ijk} (\partial_j {\cal B}_k)$
and $- \lambda \dot \rho = \lambda \Box \bar C_0$. As a consequence, we have
\begin{eqnarray}
s_b Q_d &=& {\displaystyle \int} d^3 x  \Bigl [\epsilon^{ijk} \bigl \{ (\partial_i B_j) B_k
+ (\partial_i {\cal B}_j) {\cal B}_k \bigr \} \nonumber\\
&-& (\partial^0 C^i - \partial^i C^0) \partial_i \rho
+ \lambda (\partial_0 \partial_0 \bar C_0 - \partial_i \partial_i \bar C_0)  \Bigr ].
\end{eqnarray}
Exploiting $\partial_0 \bar C_0 = - 2 \rho + \partial_i \bar C_i$ and performing
a partial integration, it can be shown that the following identity is true, namely;
\begin{eqnarray}
- {\displaystyle \int} d^3 x  \lambda \dot \rho = 
- 2 {\displaystyle \int} d^3 x  \lambda \dot \rho + {\displaystyle \int} d^3 x
(\partial^0 \bar C^i - \partial^i \bar C^0) \partial_i \lambda. 
\end{eqnarray}
As a result of the above equality, it is clear that
\begin{eqnarray}
s_b Q_d &=& - i \bigl \{Q_d, Q_b \bigr \} =
{\displaystyle \int} d^3 x  \Bigl [\epsilon^{ijk} \bigl \{ (\partial_i B_j) B_k
+ (\partial_i {\cal B}_j) {\cal B}_k \bigr \} \nonumber\\
&-& (\partial^0 C^i - \partial^i C^0) \partial_i \rho 
+ (\partial^0 \bar C^i - \partial^i \bar C^0) \partial_i \lambda  \Bigr ] \equiv W,
 \end{eqnarray}
where $W$ is defined in (46). Exactly in a similar fashion, one can show that
$s_d Q_b = - i \{ Q_b, Q_d \}$ leads to the derivation of $W$.

It can be also checked that  $s_{ab} Q_{ad} = - i \{ Q_{ad}, Q_{ab} \} = - W$. The key
steps in the above computation are as follows
\begin{eqnarray}
s_{ab} Q_{ad} &=& {\displaystyle \int} d^3 x  \Bigl [ 
(\partial^0 \bar B^i - \partial^i \bar B^0) \bar {\cal B}_i
- \epsilon^{ijk} (\partial_i \bar B_j) \bar B_k
+ \rho \dot \lambda \nonumber\\
&-& (\partial^0 \bar C^i - \partial^i \bar C^0) \partial_i \lambda \Bigr ],
\end{eqnarray}
where we have used equations (16), (21) and (35). Using the equations of motion from (24),
one can express the above equation as given below
\begin{eqnarray}
s_{ab} Q_{ad} &=& {\displaystyle \int} d^3 x  \Bigl [ 
- \epsilon^{ijk} \bigl \{ (\partial_i \bar B_j) \bar B_k + (\partial_i \bar {\cal B}_j) 
\bar {\cal B}_k \bigr \} \nonumber\\
&-& (\partial^0 \bar C^i - \partial^i \bar C^0) \partial_i \lambda +
 (\partial^0  C^i - \partial^i  C^0) \partial_i \rho \Bigr ].
\end{eqnarray}
Exploiting the constrained field equations (i.e. $B_\mu - \bar B_\mu = \partial_\mu \phi_1,
{\cal B}_\mu - \bar {\cal B}_\mu = \partial_\mu \phi_2$), we can re-express the above 
equation as follows
\begin{eqnarray}
s_{ab} Q_{ad} &=& {\displaystyle \int} d^3 x  \Bigl [ 
- \epsilon^{ijk} \bigl \{ (\partial_i  B_j)  B_k + (\partial_i  {\cal B}_j) 
 {\cal B}_k \bigr \} \nonumber\\
&-& (\partial^0 \bar C^i - \partial^i \bar C^0) \partial_i \lambda +
(\partial^0  C^i - \partial^i  C^0) \partial_i \rho \Bigr ] \equiv - W.
\end{eqnarray}
In the above, the expression for $W$ is from equation (46). In a similar fashion, it can be shown
that $s_{ad} Q_{ab} = - i \{ Q_{ab}, Q_{ad} \} = - W$.

\end{document}